\documentstyle[aps,preprint,prd,epsf,rotate]{revtex} 
\tighten 
\begin{document}
\preprint{\sf OUTP-96-11P (rev)} 
\draft
\title{Evading the Cosmological Domain Wall Problem}
\footnotetext{\sf [hep-ph/9608319]}
\author{Sebastian E. Larsson, Subir Sarkar and Peter L. White
        \vspace{5mm}}
\address{Theoretical Physics, University of Oxford, \\ 
          1 Keble Road, Oxford OX1 3NP \vspace{6mm}}
\maketitle
\begin{abstract}
Discrete symmetries are commonplace in field theoretical models but
pose a severe problem for cosmology since they lead to the formation
of domain walls during spontaneous symmetry breaking in the early
universe. However if one of the vacuua is favoured over the others,
either energetically, or because of initial conditions, it will
eventually come to dominate the universe. Using numerical methods, we
study the evolution of the domain wall network for a variety of field
configurations in two and three dimensions and quantify the rate at
which the walls disappear. Good agreement is found with a recent
analytic estimate of the termination of the scaling regime of the wall
network. We conclude that there is no domain wall problem in the
post-inflationary universe for a weakly coupled field which is not in
thermal equilibrium.
\end{abstract}
 \widetext

\section{Introduction}

It was first noted by Zel'dovich, Kobzarev and Okun \cite{zko75} that
the restoration of spontaneously broken {\em discrete} symmetries at
high temperatures in the early universe poses severe problems for its
subsequent evolution.\footnote{It has been argued recently
\cite{noproblem}, following earlier work \cite{nonrest}, that symmetry
restoration may not neccessarily occur, particularly if the scalar
field has no gauge interactions. Whether this is still true when
non-perturbative effects are taken into account is currently under
investigation \cite{nonpert}.} If the manifold ${\cal M}$ of
degenerate vacua of the theory is disconnected such that the homotopy
group $\pi_{0}({\cal M})$ is non-trivial, then sheet-like topological
defects --- domain walls --- form at the boundaries of the different
degenerate vacuua during the symmetry breaking phase transition, due
to the existence of a causal particle horizon in a decelerating
universe \cite{kib76}. The subsequent evolution of the wall network
can be studied using the techniques of percolation theory \cite{sta79}
and is such that the energy density in the walls eventually comes to
dominate the total energy density. The consequence is an inflationary
phase which suffers from a `graceful exit' problem in that it creates
an universe devoid of any matter \cite{sec86,reviews}. This can be
avoided if the energy scale associated with the discrete symmetry
breaking is low enough; however in order to avoid generating excessive
anisotropy in the cosmic microwave background, the energy scale is
further restricted to be smaller than $\sim1$~MeV \cite{reviews}. Thus
walls formed at any higher energy scale must be {\em unstable} and
have decayed away long before the present epoch.

Attempts to introduce instability in domain walls which are expected
in field theoretical models have usually focussed on the role of
non-renormalizable operators in the Lagrangian which may arise due to
violation of global symmetries by Planck-scale gravitational effects
\cite{grav}. These can `tilt' the potential so as to favour one
particular minimum which comes to dominate the universe because the
{\em pressure} of the favoured vacuum eventually wins out over the
restraining tension of the domain walls \cite{zko75,pressure}. An
example from particle physics is the addition of a dimension-5
operator to the Lagrangian of the next-to-minimal supersymmetric
standard model, in which the usual supersymmetric Higgs sector is
supplemented by a gauge singlet superfield. Here $Z_3$ domain walls
are expected to be created in the Higgs fields during the electroweak
symmetry breaking phase transition at $T{\sim}m_{W}$
\cite{nmssm}. However the dissipation of the wall network in this case
releases the contained energy as high energy interacting particles
which can interefere adversely with cosmological processes such as
primordial nucleosynthesis \cite{bbnrev}, hence the walls must
disappear well {\em before} this epoch \cite{asw95}. In order to
quantify this constraint, it is neccessary to study the rate at which
the wall network dissipates for a given pressure difference between
the different vacuua.

Another way in which the domain wall network may be destabilized is
through a suitable choice of initial conditions, viz. if the
probability distribution has a {\em bias} for one vacuum over the
others. This can be due to a prior non-equilibrium phase transition,
for example cosmological inflation, which can displace weakly coupled
light fields from the minima of their potential \cite{inflfluc},
leading to a biased initial state \cite{lto93}. It has been suggested
that the resulting unstable domain walls in a light scalar field may
be relevant for the formation of large-scale structure in the universe
\cite{structure}. Recently, both an analytic \cite{hin96} and a
numerical \cite{clo96} study of such biased domain walls have been
carried out.

In this paper we make a systematic investigation of domain wall
network evolution for both cases, viz. `pressure' (\S III (A)) and
`bias' (\S III (B)) and and compare our results with simple analytic
estimates of the rate at which the walls disappear, as well as with
previous numerical work \cite{hin96,clo96}. We also study (\S III (C))
the effects of biased initial field configurations resulting from a
period of primordial inflation. We conclude (\S IV) with a discussion
of the physical implications of our results.

\section{Numerical Techniques}

The generic potential which exhibits the problem under discussion is
\begin{equation}
 V (\phi) = V_0 \left(\frac{\phi^2}{\phi_0^2}-1\right)^2 .
\end{equation}
This has two degenerate vacuua, $\phi=\pm\phi_0$ separated by a
potential barrier $V_0$, so that the spontaneous breaking of the $Z_2$
symmetry when the universe cools below the temperature $T\sim\,V_0$
will result in the formation of domain walls. We wish to simulate the
evolution of the resulting wall network by solving the field equations
on a specified lattice. The basic problem in such numerical studies is
that there are two very different length scales involved, viz. the
wall width,
\begin{equation}
 w_0 \sim \frac{\phi_0}{\sqrt{V_0}}\ ,
\end{equation}
and the size of the simulation box. The first is constant in physical
coordinates whilst the second must be large relative to a typical
domain size, which scales with the expansion of the Universe. However
the wall thickness is in general much smaller than the size of of the
wall network so one can sensibly assume that the walls behave like
two-dimensional relativistic membranes, given that their internal
structure is expected to be of negligible importance for the
evolution. Two routes have so far been pursued in applying this idea
to computer simulations of domain wall dynamics.

Kawano \cite{kaw90} considered an effective action obtained by
expanding in $w_0/R$, where $R$ is the radius of curvature of the
wall, and retaining only the zeroth order term. The resultant
Nambu-type action yields \cite{is84} the evolution equation
\begin{equation}
 \ddot{R} + 3 \frac{\dot{a}}{a} \dot{R} (1-\dot{R}^2)
  = -\frac{2(1-\dot{R}^2)}{R}\ ,
\end{equation}
where $a$ is the cosmological scale-factor of the Robertson-Walker
metric for an Einstein-DeSitter universe, $g_{\mu\nu}={\rm diag}[-1,
a(t)^2, a(t)^2, a(t)^2]$. Unfortunately this otherwise attractive
approach leads to severe numerical instabilities (of the type
discussed in a slightly different context \cite{at89}), hence cannot
be fruitfully pursued.

The second approach, which we use in the present work is due to Press,
Ryden and Spergel \cite{prs,prs90}. Consider the classical equation of
motion for the Higgs field in an expanding background:
\begin{equation}
 \frac{\partial^2 \phi}{\partial \eta^2} + 2 \frac{d \ln a}{d
  \ln \eta} \frac{1}{\eta} \frac{\partial \phi}{\partial \eta} -
  \nabla^2 \phi = - a^2 \frac{\partial V}{\partial \phi}\ ,
\end{equation}
where $\eta$ is the conformal time (${\rm d}\eta\equiv{\rm d}t/a(t)$)
which measures the {\em comoving} distance traversed by light since
the big bang. Here the numerical problem is manifest in the $a^2$ term
on the rhs which makes the potential barrier appear higher as time
goes on, resulting in a wall solution which grows increasingly narrow
(in comoving cordinates) with time. In attempting to evolve an
equation of this type directly one would find that the wall solutions
in the $\phi$ field quickly become so narrow as to be
unresolvable. The solution is to generalize the equation of motion to:
\begin{equation}
 \frac{\partial^2 \phi}{\partial \eta^2} + \alpha \frac{{\rm d} 
  \ln a}{{\rm d} \ln \eta} \frac{1}{\eta} \frac{\partial\phi}{\partial\eta} -
  \nabla^2\phi = - a^\beta\frac{\partial V}{\partial\phi}\ ,
\label{eqn}
\end{equation}
and then set $\beta=0$ in order to freeze the wall size in comoving
coordinates. We also set $\alpha=3$ to ensure momentum conservation since
this requires that we have \cite{prs} 
\begin{equation}
 \alpha + \frac{\beta}{2} = 3\ .
\label{alpha}
\end{equation}
Press {\em et al.} have discussed in some detail the justification for
this approximation and demonstrated that it has a negligible effect on
the evolution by performing simulations comparing $\alpha=\beta=2$ and
$\alpha=3, \beta=0$ results over a limited range of $\eta$. We perform
a test of a complementary nature as described below which supports
their contention that the numerical method is robust and unlikely to
give spurious results.
 
Following ref.\cite{prs} we implement this procedure using the
standard finite differenceing scheme embodied in the expressions
\begin{eqnarray}
	& \delta \equiv \frac{1}{2} \alpha \frac{\Delta\eta}{\eta} 
	 \frac{{\rm d}\ln a}{{\rm d} \ln \eta}\ , & \\
	& ({\nabla}^2 \phi)_{i,j,k} \equiv \phi_{i+1,j,k} + \phi_{i-1,j,k} +
         \phi_{i,j+1,k} + \phi_{i,j-1,k} + \phi_{i,j,k+1} + \phi_{i,j,k-1} -
         6\phi_{i,j,k}\ , & \\
	& \dot{\phi}^{n+1/2}_{ijk} =
         \frac{(1-\delta)\dot{\phi}^{n-0.5}_{ijk} + \Delta\eta ({\nabla}^2 
         \phi^{n}_{ijk} - \partial V/\partial\phi^{n}_{ijk})}{1+\delta}\ , & \\
	& \phi^{n+1}_{ijk} = \phi^{n}_{ijk} +  \Delta \eta
         \dot{\phi}^{n+0.5}_{ijk} . &
\end{eqnarray}
We also use the same algorithm \cite{prs} for measuring the total wall
area in the $\phi$ field. Our simulations were run in both two and
three dimensions, on a $128{\times}128{\times}128$ grid ($D=3$) and a
$1024{\times}1024$ grid ($D=2$).

\section{Wall Network Evolution}

The general evolution of the domain wall network is well known
\cite{reviews}; the most interesting phase is that of `Kibble
scaling', in which the walls have a correlation length $\xi$ in
comoving units given by
\begin{equation}
 \xi \simeq v \eta\ ,
\label{v}
\end{equation}
where $v$ is the velocity of wall propagation. The simplest variable
which tracks the evolution is then the comoving area of the wall
network per unit volume, $(A/V)$, which scales as
\begin{equation} 
 (A/V) \simeq (A/V)_{0}\frac{\xi_0}{\xi}
         \simeq \frac{\xi_0}{v(\eta-\eta_0)}\ ,
\label{scaling}
\end{equation}
while the physical energy density contained in walls behaves as
$a^{-1}$ times this. We find that there is indeed a scaling region
with
\begin{equation} 
 (A/V) \propto \eta^{-\nu}, \quad \nu = \left\{ \begin{array}{ll} 
       0.95 \pm 0.08 & \qquad \mbox{for $D = 2$\ ,} \nonumber \\
       0.92 \pm 0.08 & \qquad \mbox{for $D = 3$\ ,}
       \end{array} \right.
\label{nu}
\end{equation}
in agreement both with Eq.(\ref{scaling}) and other simulations
\cite{prs,clo96}. (The errors quoted here are purely
statistical). Hence we can calculate $v$, as defined in Eq.(\ref{v}),
to be
\begin{equation} 
 v = \left\{ \begin{array}{ll}
     0.7 \pm 0.1 & \qquad \mbox{for $D = 2$\ ,} \nonumber \\
     0.5 \pm 0.1 & \qquad \mbox{for $D = 3$\ ,}
     \end{array} \right.
\end{equation}
(Note that the velocity which was found previously to be $\sim0.4$ in
three dimensions \cite{prs}, was somewhat differently defined.) We
shall henceforth simplify our algebra by normalising the values of the
comoving energy density and scale factor to be unity at the initial
time of our simulation, here denoted with the subscript $0$. The
evolution of the network thus proceeds as follows. For the initial few
timesteps, with $\eta\lesssim10$, the correlation length is less than
the wall thickness, before the network enters the scaling regime of
equation~(\ref{scaling}) which ends at $\eta\sim100$ when it becomes
comparable to the size of the lattice.

\begin{figure}[htb]
\epsfxsize=12cm
\centerline{\epsfbox{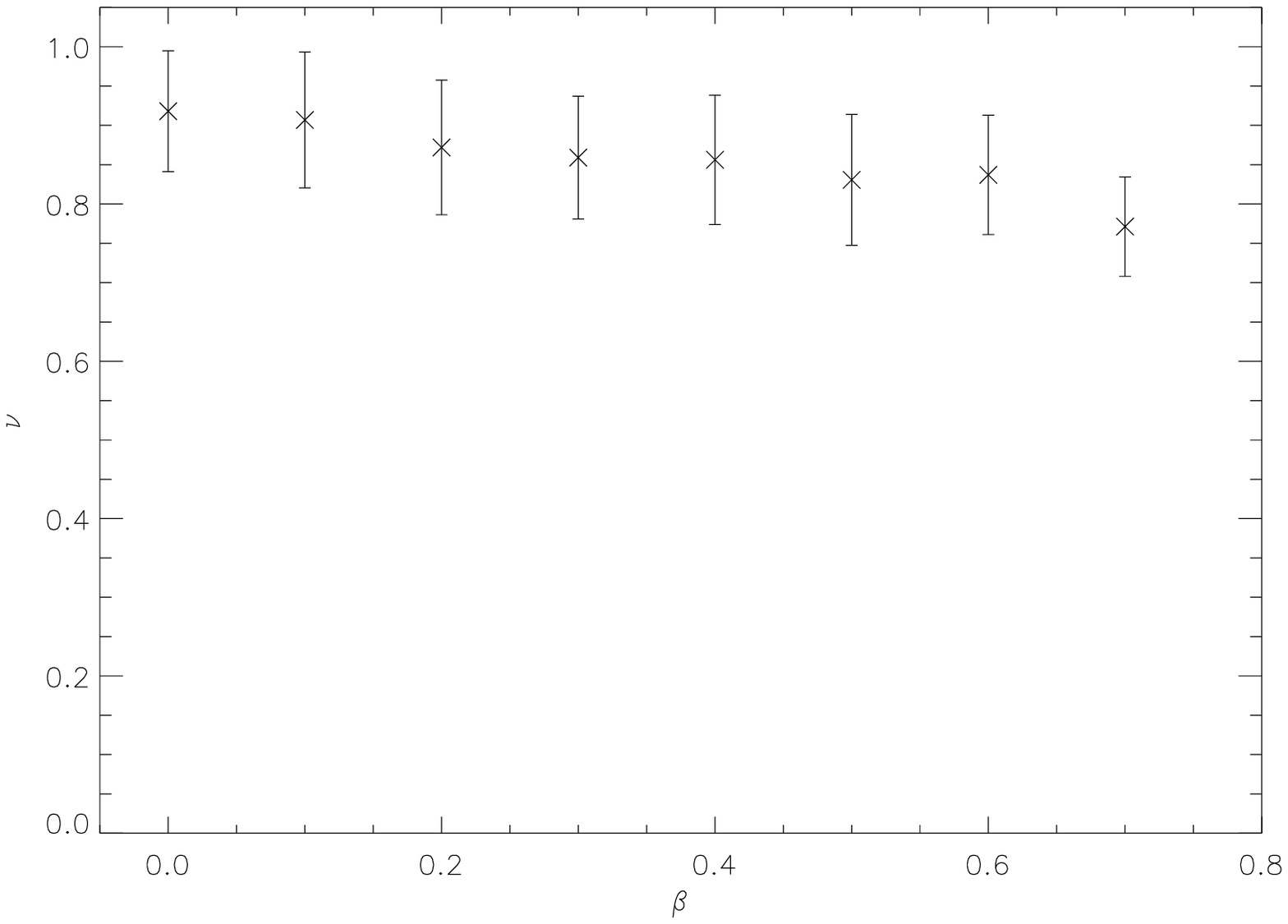}}
\caption{A test of the numerical approximation used. The scaling index
$\nu$ of the wall network is seen not to vary significantly as the
parameter $\beta$ in the equation of motion is varied.}
\label{checkevol}
\end{figure}

To verify the suitability of the numerical method used we ran a suite
of simulations for $D=3$ where $\beta$ is increased in steps of 0.1
from zero and $\alpha$ is adjusted according to Eq.(\ref{alpha}). In
order for the domain wall to be resolvable, it should occupy at least
one (preferably several) lattice spacing(s); this corresponds to a
limiting value of $\beta\simeq0.66$ in our simulations. In
figure~\ref{checkevol} we show that the exponent $\nu$ in
Eq.(\ref{nu}) does not change significantly as $\beta$ is increased up
to this value.\footnote{There appears to be a small systematic trend
of decreasing $\nu$ with increasing $\beta$. Although not
statistically significant, this may just reflect the increasing
efficiency of the area measuring algorithm at picking up small bubbles
at late times, as $\beta$ is increased.} Overall the gross features of
the evolution appear to be insensitive to the numerical approximation
used; in particular the onset and termination of the scaling regime
are {\em not} dependent on $\beta$. This adds further weight to the
conclusion of Press {\em et al.} that there is no difference between
the dynamics of `real' walls which have constant physical thickness,
and the `artificial' walls in these simulations which have constant
comoving thickness.

\subsection{Pressure}

\begin{figure}[htb]
\epsfxsize=12cm
\centerline{\epsfbox{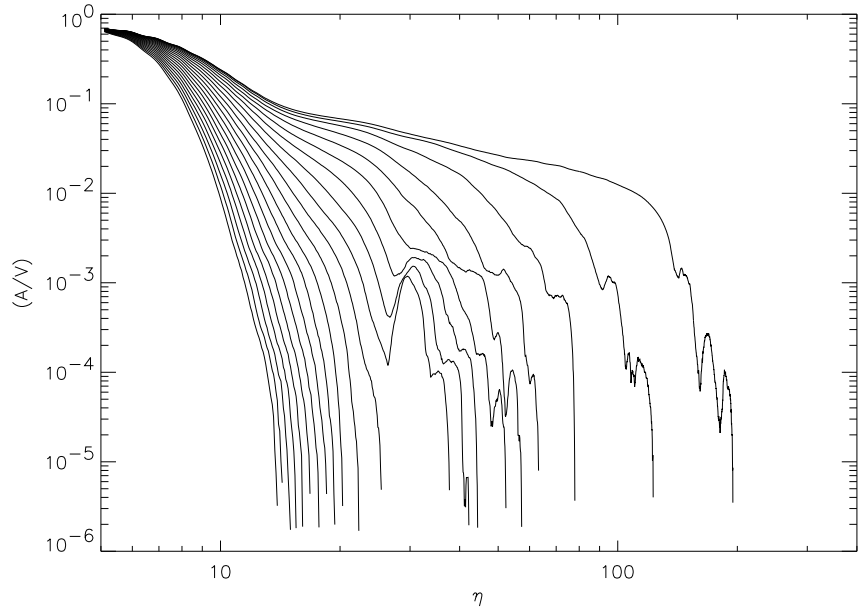}}
\caption{Comoving area against conformal time in 3 dimensions, with 
 the pressure $\mu$ in the range $0-0.2$.}
\label{pressevol3D}
\end{figure}

\begin{figure}[htb]
\epsfxsize=12cm	
\centerline{\epsfbox{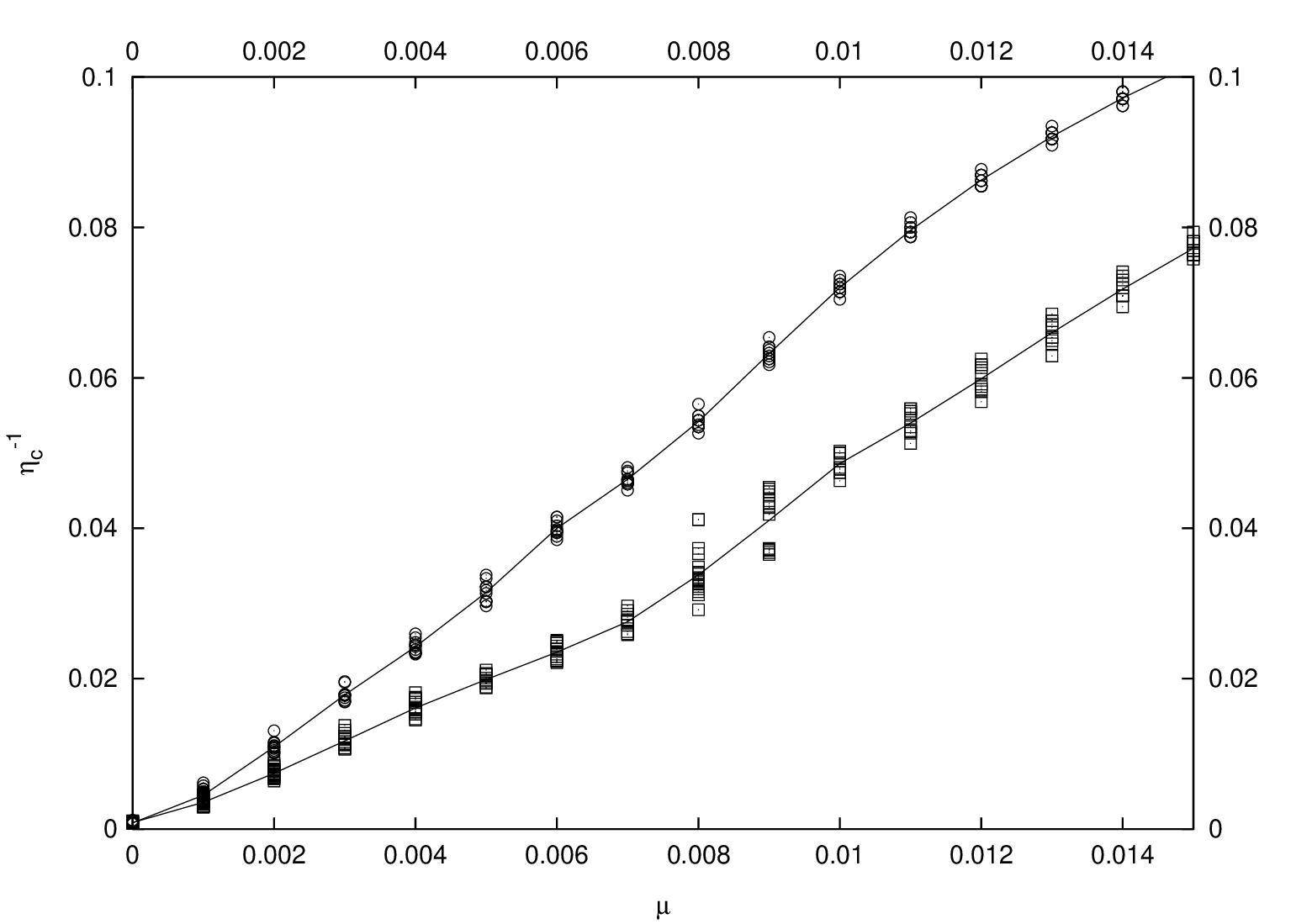}}
\caption{$\eta_{\rm c}^{-1}$ against pressure $\mu$ in 2 dimensions,
 where $\eta_{\rm c}$ is the conformal time at which the product of
 $\eta$ and the comoving area density has fallen by a factor
 $10$ (circles) or $100$ (squares). The line connects the mean
 value of $\eta_{\rm c}$ over all the runs.}
\label{press2D}
\end{figure}

\begin{figure}[htb]
\epsfxsize=12cm
\centerline{\epsfbox{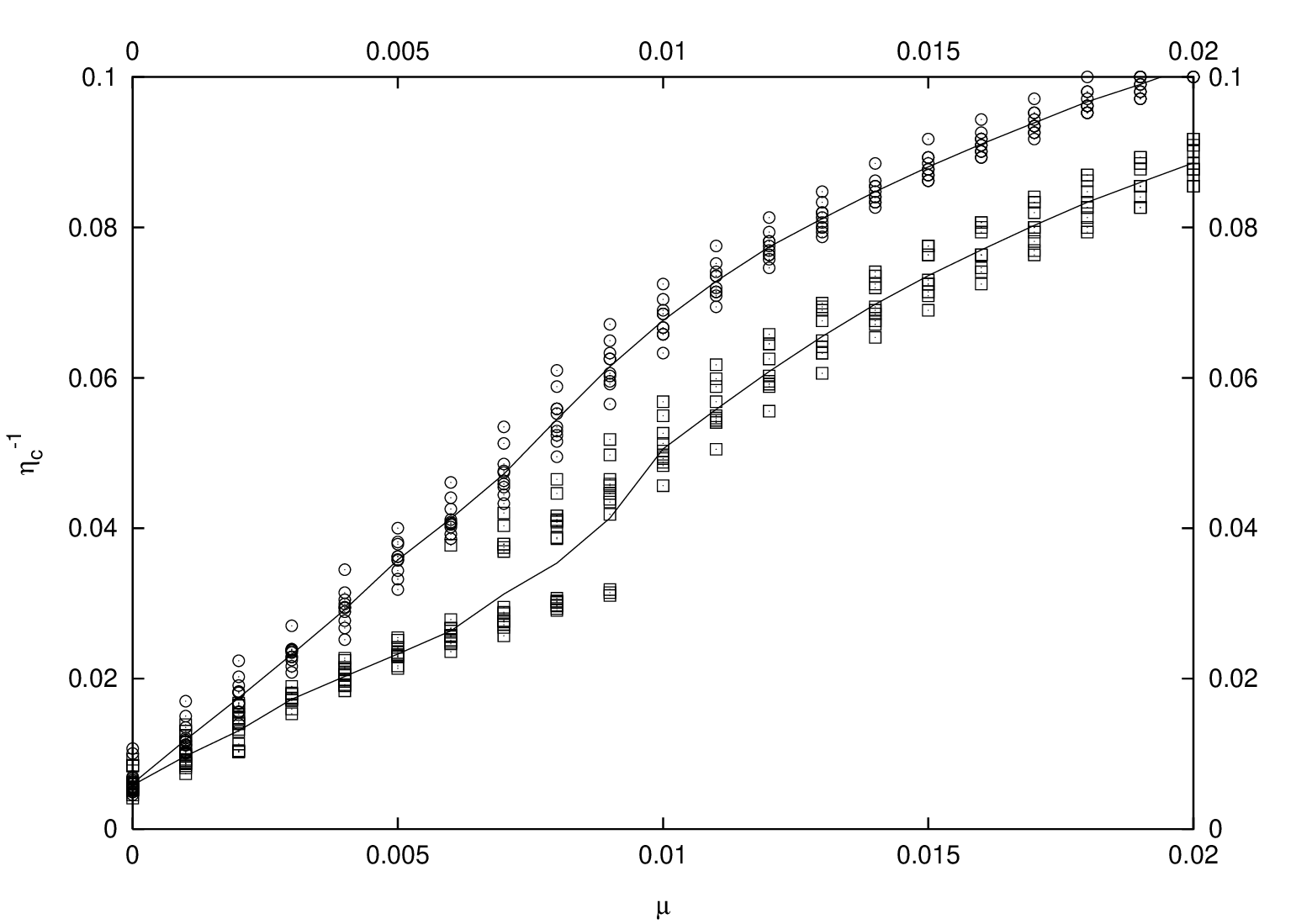}}
\caption{$\eta_{\rm c}^{-1}$ against pressure $\mu$ in 3 dimensions,
 where $\eta_{\rm c}$ is the conformal time at which the product of
 $\eta$ and the comoving area density has fallen by a factor
 $10$ (circles) or $100$ (squares). The line connects the mean
 value of $\eta_{\rm c}$ over all the runs.}
\label{press3D}
\end{figure}

A well known way \cite{pressure,solitons} to evade the causal limit on
the disappearance of the domain walls is to include a pressure term in
the potential, viz.
\begin{equation}
 V (\phi) = V_0 \left[\left(\frac{\phi^2}{\phi_0^2}-1\right)^2
            + \mu \frac{\phi}{\phi_0}\right] ,
\label{potpress}
\end{equation}
The dynamics is now expected to be dominated by two competing forces,
the surface tension $\sigma/R$ where $R\simeq\,a\xi$ is the radius of
curvature of the wall structure, and the pressure which is equal to
the differences in energy density of the two minima of the potential,
$\delta{V}$. Then we expect simple scaling behaviour (\ref{scaling})
until some conformal time $\eta_{\rm c}$ at which the pressure becomes
comparable to the surface tension, and the wall network disappears
exponentially fast. Now $\eta_{\rm c}$ is given by
\begin{equation}
 \sigma \frac{a(\eta_0)}{(\eta_{\rm c}-\eta_0) a(\eta_{\rm c})} = \delta {V}\ ,
\end{equation}
where the subscript $0$ indicates the value at the begining of the
evolution. In the present case we are scaling the fields with time so
as to remove factors of $a$ and set $\eta_0=1$ so this equation will
reduce simply to
\begin{equation}
 \frac{\sigma}{\eta_{\rm c}} = 2 \mu\ .
\label{presseqn}
\end{equation}
We display this behaviour in figure~\ref{pressevol3D}, where the
comoving area density is plotted against conformal time for several
choices of $\varepsilon$. As expected, there is an exponential
fall-off from the scaling regime at some value of $\eta_{\rm c}$ which
decreases with increasing $\mu$. The `bounces' in the curves occur as
the bubbles of the disfavoured vacuum collapse and radiate away the
energy contained in the walls in the form of Goldstone bosons; as
noted earlier \cite{wid89} a large fraction of the energy is lost in a
few bounces. The relation between $\mu$ and $\eta_{\rm c}$ is shown in
figures~\ref{press2D} and \ref{press3D}, which show excellent
agreement with the expected behaviour (\ref{presseqn}) to within a
factor of 2. (Here $\eta_c$ is defined as the value of $\eta$ at which
the product $\eta(A/V)$ has fallen off by some factorm taken here to
be 10 and 100 for illustration.) The actual rate of the decay of the
wall network after the exponential decay has set in is hard to
measure. We find here that assuming a behaviour of form $(A/V)_{\rm
c}\sim\eta^{-1}\exp[-\kappa(\mu\eta)^n]$ with $\kappa$ some constant,
$n$ is $\approx 2\pm1$ ($D=2$) and $\approx3\pm1$ ($D=3$).

The implication of this is that we can regard the walls as simply
disappearing as a consequence of the difference in energy between the
two minima of the potential $\delta{V}$ at a correlation length
$R_{\rm c}=\sigma/\delta{V}$ (in physical, not comoving, units), where
again the constant of proportionality is around unity. Hence, if we
have walls forming at the electroweak scale and we require that they
disappear before nucleosynthesis \cite{bbnrev}, we find
\begin{equation}
 \frac{\delta{V}}{\sigma} > \frac{1}{10^{10}\ {\rm cm}}
  \sim 10^{-24}\ {\rm GeV}\ ,
\end{equation}
in good agreement with previous estimates made from physical arguments
as to the time of pressure domination \cite{pressure} and
2-dimensional thin wall simulations \cite{aw95}.

\subsection{Bias}

Next we consider the possibility of a deviation from scaling behaviour
due to a {\em biased} initial probability distribution. For
concreteness we consider the $Z_2$ case where there are only two
distinct vacuua and we generate the initial configuration with a
probability
\begin{equation}
 p_+ = 0.5 + \varepsilon 
\label{varepsilon}
\end{equation}
that each initial domain is in the $+$ phase, where $\varepsilon>0$. A
similar exercise has been performed recently by Coulson, Lalak and
Ovrut \cite{clo96} who find that the domain wall network then evolves
much more rapidly than in the usual case of Eq.(\ref{scaling}), and
the favoured domain rapidly dominates the universe. Such a scenario is
interesting because a bias in a light scalar field which is not in
thermal equilibrium can be generated naturally by a previous epoch of
inflation. Other possible ways in which a scalar field is more likely
to find itself in one minimum than the other are if the symmetry is
approximate, so that the two minima are not exactly degenerate, or if
the symmetry breaking occurs through some intermediate phase which
allows one minimum to be preferentially populated.

\begin{figure}[htb]
\epsfxsize=12cm	
\centerline{\epsfbox{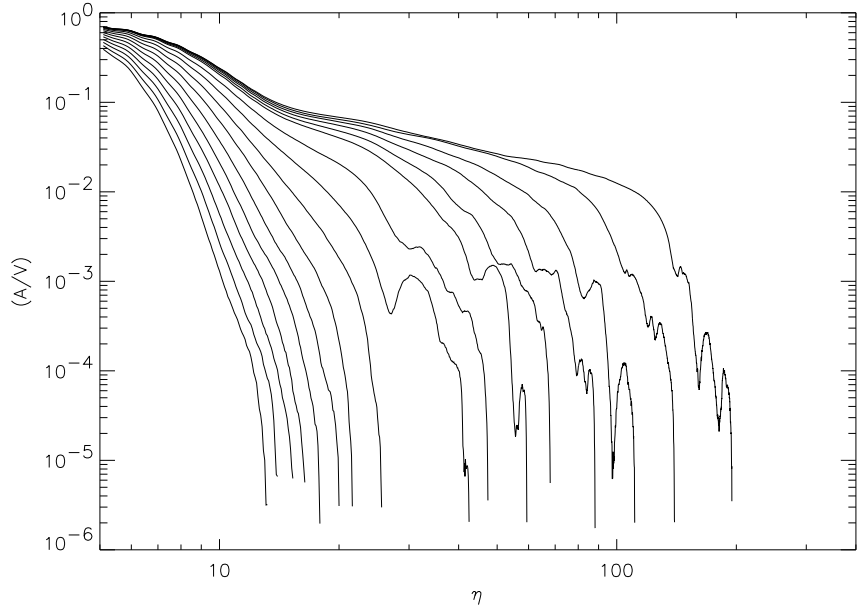}}
\caption{Comoving area against conformal time in 3 dimensions with the
 bias $\varepsilon$ in the range $0-0.03$.}
\label{biasevol3D}
\end{figure}

The evolution is shown in figure~\ref{biasevol3D}, where we plot
comoving area density against conformal time for several choices of
$\varepsilon$, ranging upto $0.03$. Typically, the behaviour is similar
to the usual case of no bias, with an exponential decay of the
comoving energy density starting at some critical timescale $\eta_{\rm
c}$. (For $\eta\lesssim10$ the behaviour is unphysical since $\xi$ is
less than the domain wall thickness (here scaled to be 5), while for
$\eta\gtrsim100$, $\xi$ is approaching the physical size of the box,
and hence the behaviour is again unphysical.) Notice that the wall
network dissipates rapidly as $\varepsilon$ is increased above zero,
as was also found in ref.\cite{clo96}. (Again we see the `bounces'
associated with the radiating away of the energy contained in the wall
network.)

Hindmarsh \cite{hin96} has recently made an analytic study of biased
domain walls. He considers the evolution of the domain walls in terms
of a scalar field constructed so that its zeros correspond to the
locations of the domain walls, and whose evolution can be calculated
in terms of Gaussian average field configurations; the expectation for
the wall surface area density in $D$ dimensions is
\begin{equation}
 (A/V) \sim \frac{1}{\eta} \exp\left(-\kappa\varepsilon^2\eta^D\right ) ,
\label{markh}
\end{equation}
where $\kappa$ is some constant which should, in our units, be of
order unity. The above result was obtained using rather sophisticated
techniques but we can derive it from a much simpler (but perhaps less
trustworthy) counting argument. We consider the wall evolution to
proceed in such a way that the universe naturally divides itself up
into domains of size $\xi{\simeq}v\eta$ (in comoving
coordinates). Such domains have been causally connected, and so have
had time enough to organize themselves to be either all $+$ or all
$-$. Hence we can see that the comoving area density in the absence of
bias will behave as $(A/V)\simeq\xi^{-1}\simeq(v\eta)^{-1}$ as
expected.  We can now examine the behaviour of the comoving area
density in the more complicated situation when there is a bias. Now we
expect a domain of conformal size $\xi$ to become a $+$ domain if most
of its $\xi^3$ subdomains are $+$, and a $-$ domain otherwise. (We
have normalized so that the size of the initial domains is unity.) If
the probability of each subdomain entering the positive minimum is
$p_+$, the probability for the whole domain to be $-$ is given by a
sum over values of a Poisson distribution which can, in any
interesting case, be taken to be a Gaussian integral, viz.
\begin{equation}
 {\cal P} (\hbox{\rm domain of $N$ sites, mostly $-$}) = \hbox{\rm erf}
  \left(\frac{\bar{N} - N/2}{\sigma_N}\right )
 = \hbox{\rm erf} (\sqrt{2}\varepsilon\sqrt{N})\ ,
\label{bigproby}
\end{equation}
where $\bar{N}=(0.5+\varepsilon)N$ is the mean number of $+$
sites and $\sigma_N=\sqrt{\bar{N}}$ is the standard deviation. Here
the Gaussian integral ${\rm erf}(x)\equiv\int_{x}^\infty\,{\rm d}t\,{\rm
e}^{-t^2/2}/\sqrt{2\pi}$ can be adequately approximated for large $x$
by
\begin{equation}
 \hbox{\rm erf} (x) = \frac{1}{x\sqrt{2\pi}} {\rm e}^{-x^2/2} .
\end{equation}
We see that for a box with $\xi$ sites, the probability given in
Eq.(\ref{bigproby}) is just ${\rm erf}(\sqrt{2}\varepsilon\xi^{D/2})$
and hence we expect exactly the exponential behaviour shown earlier in
Eq.(\ref{markh}). The extra factor of $(\xi^{D/2})^{-1}$ is cancelled
by a combinatoric factor from the many different ways in which we can
lay down our domains upon the overall network. This uses the fact that
a cluster containing $N$ sites will typically scale as $N^{1/2}$,
using the results of percolation theory as recently applied to similar
problems in cosmology \cite{lot95}.

\begin{figure}[htb]
\epsfxsize=12cm 
\centerline{\epsfbox{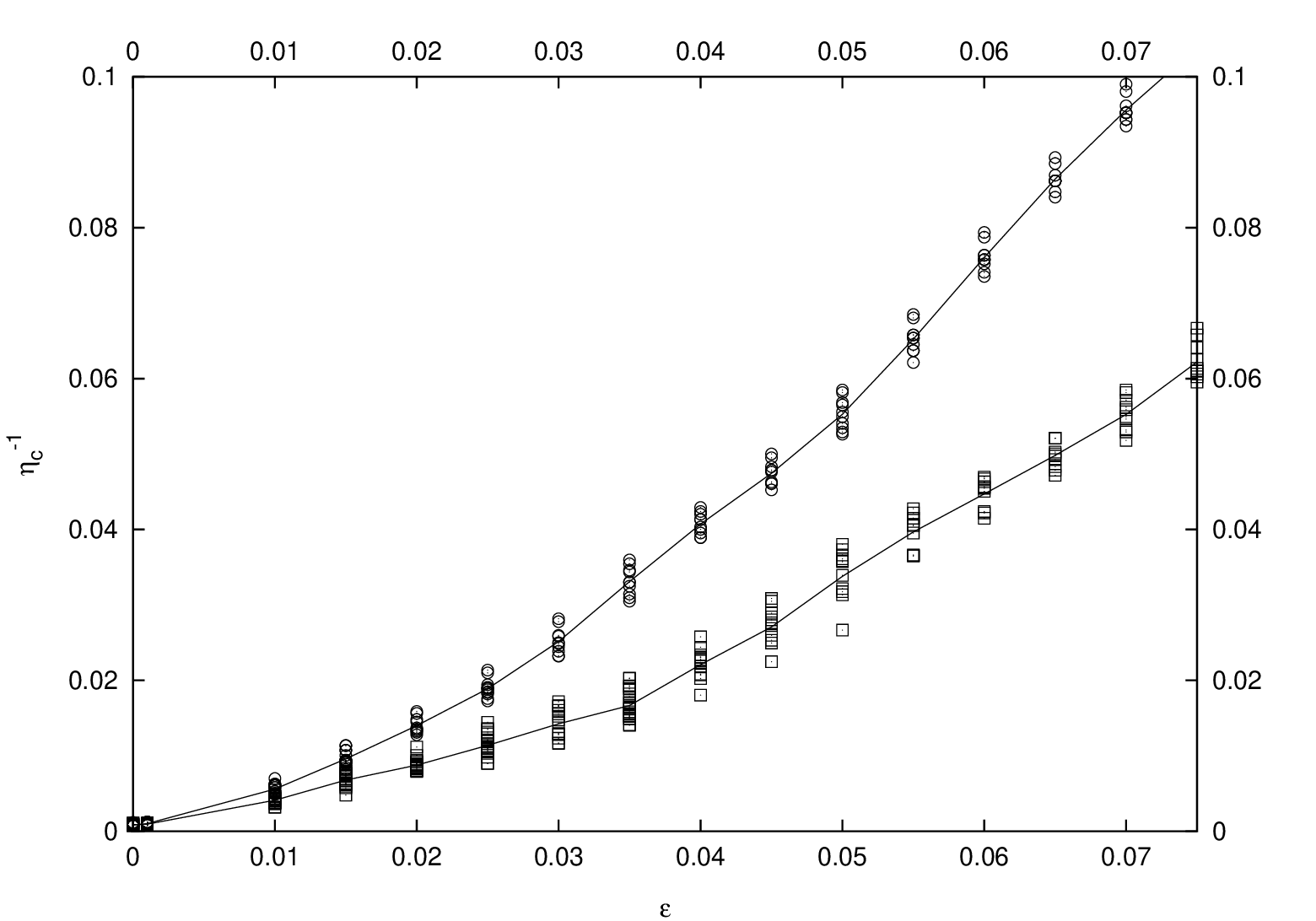}}
\caption{$\eta_{\rm c}^{-1}$ against bias $\varepsilon$ in 2
 dimensions, where $\eta_{\rm c}$ is the conformal time at which the
 product of $\eta$ and the comoving area density has fallen by a
 factor $10$ (circles) or $100$ (squares). The line connects
 the mean value of $\eta_{\rm c}$ over all the runs.}
\label{bias2D}
\end{figure}

\begin{figure}[htb]
\epsfxsize=12cm
\centerline{\epsfbox{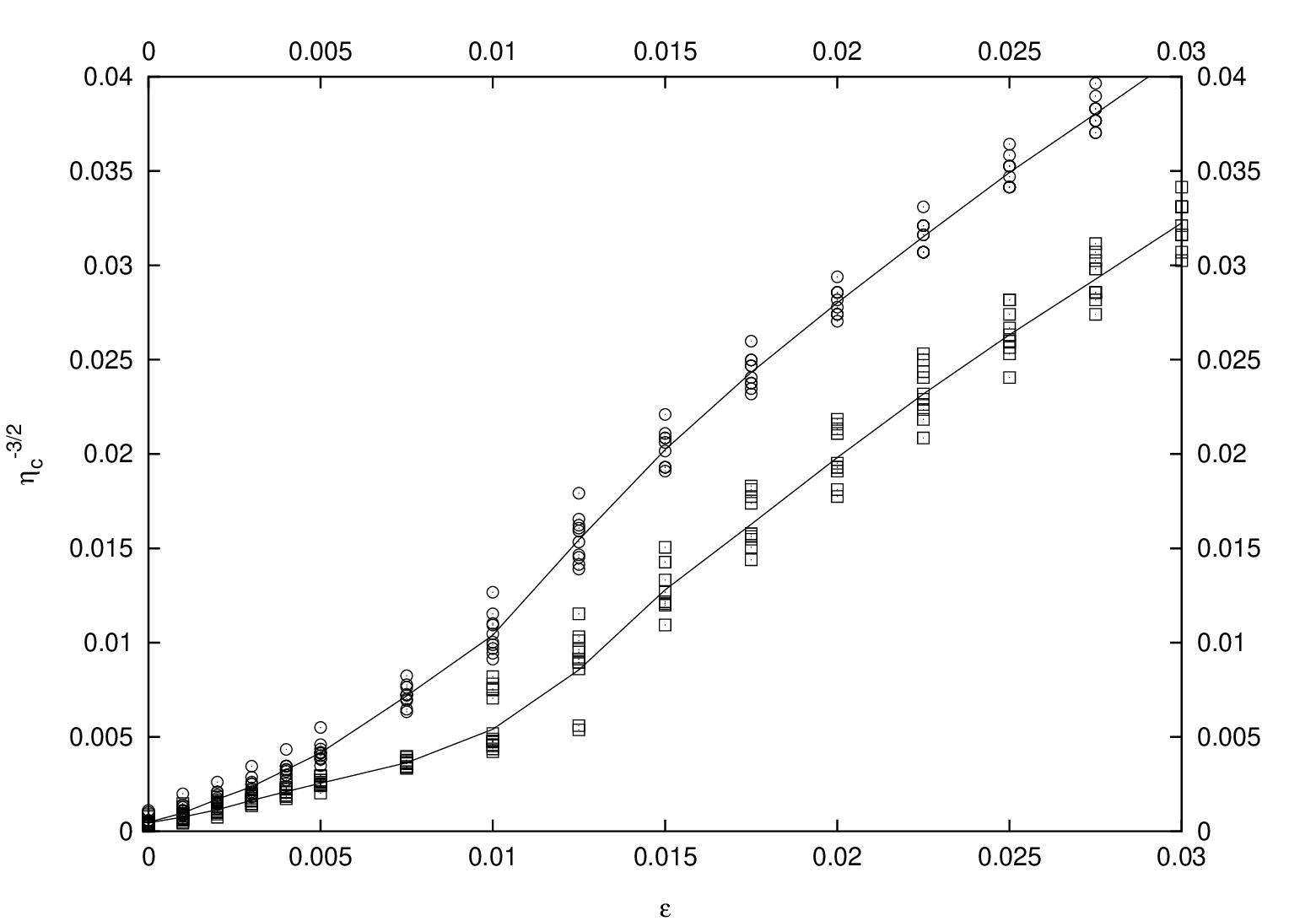}}
\caption{$\eta_{\rm c}^{-1}$ against bias $\varepsilon$ in 3
 dimensions, where $\eta_{\rm c}$ is the conformal time at which the
 product of $\eta$ and the comoving area density has fallen by a
 factor $10$ (circles) or $100$ (squares). The line connects
 the mean value of $\eta_{\rm c}$ over all the runs.}
\label{bias3D}
\end{figure}

To verify this behaviour, we define $\eta_{\rm c}$ to be the conformal
time at which the product $\eta(A/V)$ has decreased by some
substantial factor, which we choose to be 10 and 100 for
illustration. The values of $\eta_{\rm c}^{-D/2}$ are plotted against
$\varepsilon$ for both 2 and 3 dimensions in figures \ref{bias2D} and
\ref{bias3D} and we find good agreement with the theoretical
prediction of a straight line. Note that for both cases, we have cut
off the figures at $\eta\simeq10$ since before this conformal time the
domain wall network is not yet fully formed. We also expect unreliable
results when $\eta$ reaches 100 or 1000 in three and two dimensions
respectively, since then the correlation length is approaching the
size of our whole simulation. While we can measure the relationship
between $\eta_{\rm c}$ and the bias $\varepsilon$, it is again harder
to measure the actual rate of the exponential fall off, i.e. the
exponent of $\eta$ in the exponential. We find this exponent to be
$\approx1-2$ ($D=2$) and $\approx2-3$ ($D=3$), in acceptable agreement
with theory.

Therefore we can, for cosmological purposes, regard the wall evolution
as being in the Kibble scaling regime for early times, and modelled
subsequently by an exponential fall off at some conformal time
$\eta_{\rm c}$, which will lead to an almost immediate collapse of the
wall network. The required value of $\varepsilon$ is readily
calculable in three dimensions from
\begin{equation}
\varepsilon = \left(\frac{\eta_{\rm c}}{\eta_0}\right)^{-3/2}
            = \left(\frac{T_0}{T_{\rm c}}\right)^{-3/2} ,
\end{equation}
where $\eta_{\rm c}$ ($T_{\rm c}$) is the scale factor (temperature)
at the time of disappearance and $\eta_0$ ($T_0$) at the time of
domain formation. (Note that here we put $\kappa\simeq 1$ from our
simulation result.) For example, in the event of wall formation at the
weak scale of order 100 GeV, with the requirement that the walls
disappear before the nucleosynthesis era which starts at $\sim1$ MeV
\cite{bbnrev}, we find that $\eta_{\rm c}/\eta_0\sim10^5$, and hence
that the bias in the initial probability distribution must be of order
$10^{-8}$ or greater. However if the walls form subsequent to
reheating following inflation, at an energy scale of say $10^{6}$ GeV,
the bias need only be about $10^{-14}$.

\subsection{Realistic Field Configurations}

\begin{figure}[htb]
\epsfxsize=12cm
\centerline{\epsfbox{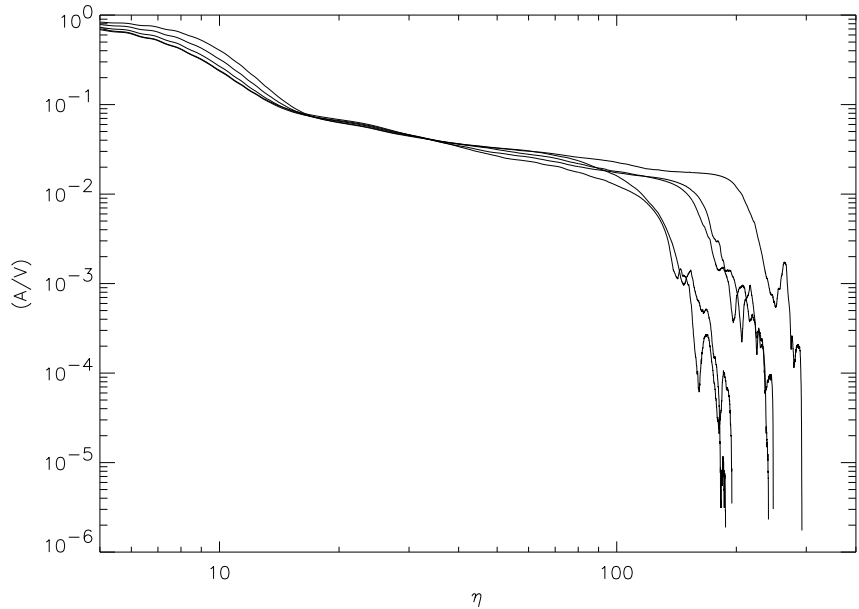}}
\caption{Comoving area against conformal time for post-inflationary
 field configurations, parametrized by the width
 $\sigma=0,0.5,1,1.5,2$ with no bias.}
\label{gauss000}
\end{figure}

\begin{figure}[htb]
\epsfxsize=12cm
\centerline{\epsfbox{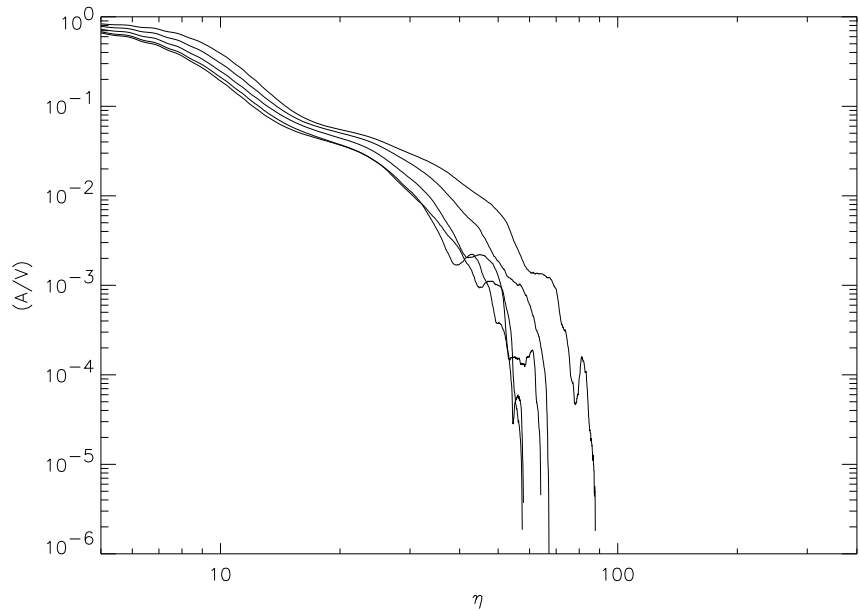}}
\caption{Comoving area against conformal time for post-inflationary
 field configurations, parametrized by the width
 $\sigma=0,0.5,1,1.5,2$ with bias $\varepsilon=0.005$.}
\label{gauss005}
\end{figure}

It is usually assumed \cite{kib76} that following cosmological
discrete symmetry breaking in a scalar field, the field values will be
uncorrelated on scales larger than the causal (particle) horizon, so
the degenerate vacuua will be populated with equally probability.
However, a prior period of cosmological inflation {\em can} result in
correlations on (apparently) super-horizon scales if the field is
sufficiently weakly coupled so as not to be in thermal
equilibrium. This is because quantum fluctuations during inflation
(with Hubble parameter $H$) will induce long-wavelength fluctuations
in all scalar fields with mass $m\ll\,H$ on spatial scales
$k^{-1}\gg\,H^{-1}$ \cite{inflfluc}. For a minimally coupled field
with vanishing potential (e.g. a (pseudo-) goldstone boson), this
leads to the formation of a classical inhomogeneous field, with
gaussian probability distribution \cite{ll90}
\begin{equation}
 {\cal P} (\phi) 
  \propto \exp \left[-\frac{(\phi-\bar{\phi}_{k})^2}{2\sigma_{k}^2}\right]\ ,
  \qquad \sigma^2_{k} \simeq \frac{H^2}{4\pi^2} \int_{k}^{k_{c}} {\rm d}\ln k\ .
\label{gauss}
\end{equation}
where $\bar{\phi}_{k}$ is the mean value of the field averaged over a
domain of size $k^{-1}$ and $k_c$ is an effective ultraviolet cutoff
imposed by physical considerations (dependent on the nature of the
field). Since inflation blows up the scale factor exponentially fast,
this provides a natural mechanism for bias since the global ensemble
average of $\phi$ may not be realized even over regions as big as our
present universe. During the post-inflationary phase, the above
distribution (\ref{gauss}) averaged over a chosen scale thus becomes
skewed \cite{lto93}. We parametrize this for the $Z_2$ case by drawing
probabilities for populating the two vacuua from the distribution
\begin{equation}
 {\cal P} (\phi) = \frac{1}{\sqrt{2\pi} \sigma}
                 \left[(0.5 - \varepsilon) {\rm e}^{-(\phi+1)^2/2\sigma^2} +
                 (0.5 + \varepsilon) {\rm e}^{-(\phi-1)^2/2\sigma^2}\right]\ ,
\end{equation}
where $\varepsilon$ is the effective bias over the chosen
scale. (Presumably its value can be calculated in principle if the
field and inflationary parameters are specified.) In figures
\ref{gauss000} and \ref{gauss005} we show the comoving area density
plotted against conformal time for several choices of $\sigma$ with
$\varepsilon=0,\ 0.005$. Surprisingly, the exponential fall-off from the
scaling regime appears to be quite insensitive to the width of the
probability distribution. This supports the suggestion \cite{lto93}
that this is an efficient way to eliminate the domain wall network.

\section{Conclusions}

We have considered two possible ways in which a cosmological domain
wall network can be made unstable. The first of these is that of
pressure, i.e. a small breaking of the degeneracy between the
minima. The physical implications of this are well understood and we
have simply confirmed numerically the usual argument \cite{pressure}
that when the typical domain size $R$ has grown to a critical value
$R_{\rm c}$, the wall energy density decays exponentially fast, so the
network disappears essentially instantaneously. The value of $R_{\rm
c}$ is
\begin{equation}
 R_{\rm c}=\frac{\sigma}{\delta\rho}\ ,
\end{equation}
where $\sigma$ is the wall surface energy density and $\delta\rho$ is
the difference in energy density between the two minima. (There may be
an additional effect due to the bias induced at the phase transition
by the energy difference between the two minima.)

The second mechanism is that of bias, where the initial field
configuration is not symmetric between the various minima of the
potential. Here we confirm previous analytical \cite{hin96} and
numerical \cite{clo96} studies which predict an exponential decay in
the energy density with a characteristic conformal time $\eta_{\rm
c}$. This is an attractive solution to the domain wall problem if we
can find a natural mechanism to bias the initial distribution,
particularly since we find the evolution to be insensitive to the
width of the distribution. Since such a bias must appear on
(apparently) causally disconnected scales, it can only occur if either
the minima are genuinely inequivalent (and so have non-degenerate
energy densities) or else if we are in a post-inflationary phase in
which the field has correlations on super-horizon scales. This will
allow the universe to more efficiently organise itself into one
preferred minimum everywhere, thus enabling the wall network to decay
away. Now an initial field configuration with correlations on all
wavelengths can be characterized by an effective bias $\varepsilon(R)$
averaged over a box of size $R$, as high frequency modes will average
out to zero. We have seen that a region of size $R$ will decay away
exponentially after a conformal time $\eta_{\rm
c}(R)\sim\varepsilon(R)^{-D/2}$. For a radiation dominated universe,
the time $t(R)$ at which the domain will decay is then
\begin{equation}
 t (R) \sim \varepsilon(R)^{-3}\ ,
\end{equation}
which must grow more slowly than $R$ to ensure that the ordinary
causal scaling bound is exceeded. Hence for bias to provide a
realistic means of eliminating wall networks through an initially
non-equilibrium field distribution, we must have that $\varepsilon(R)$
falls off with $R$ more slowly than $R^{-1/3}$. Since the spectrum of
fluctuations from inflation is (nearly) scale-invariant, we may expect
that the bias $\varepsilon(R)$ does stay approximately constant with
$R$, and hence that this mechanism provides a way of eliminating
domain walls in fields which are sufficiently weakly coupled so as not
to be in thermal equilibrium.

\acknowledgements{S.E.L. thanks CSN and KV (Sweden), ORS and OOB
(Oxford) and the Sir Richard Stapley Educational Trust (Kent), and
S.S. thanks PPARC for support. We are grateful to Steven Abel, Mark
Hindmarsh, Zygmunt Lalak and Graham Ross for helpful discussions and
to the referee for prompting us to investigate in more detail the
numerical method employed.}

\end{document}